\documentclass{PoS}

\graphicspath{{./images/}}
\usepackage{graphbox}
\title{Autonomous RPCs for a Cosmic Ray ground array}

\ShortTitle{Autonomous RPCs for a Cosmic Ray ground array}

\author{Alberto Blanco, Nuno Carolino, Orlando Cunha, Miguel Ferreira, Paulo Fonte, Luis Lopes, Ricardo Luz, Luis Mendes, Américo Pereira, Raul Sarmento\\
        Laborat\'orio de Instrumenta\c{c}\~{a}o e F\'isica Experimental de Partículas - LIP, Portugal\\}

  \author{Pedro Assis, \speaker{Ruben Concei\c{c}\~{a}o}\thanks{Acknowledges the financial support of Funda\c{c}\~ao para a Ci\^encia e Tecnologia, FCT-Portugal and Funda\c{c}\~ao de Amparo \`a Pesquisa do Estado de S\~ao Paulo, FAPESP-Brazil.} , M\'ario Pimenta, Bernardo Tom\'e\\
        Laborat\'orio de Instrumenta\c{c}\~{a}o e F\'isica Experimental de Partículas - LIP and Instituto Superior T\'ecnico - IST, Universidade de Lisboa - UL, Portugal\\
        E-mail: \email{ruben@lip.pt}}   
        
 \author{Victor Barbosa Martins, Vitor de Souza\\
   Universidade de São Paulo, Instituto de Física de São Carlos - IFSC/USP, Brazil\\}
   
    \author{Ronald Shellard\\
Centro Brasileiro de Pesquisas Físicas - CBPF, Brazil\\}

    \author{Carola Dobrigkeit\\
Universidade Estadual de Campinas, Brazil\\}

\abstract{We report on the behaviour of Resistive Plate Chambers (RPC) developed for muon detection in ultra-high energy cosmic ray (UHECR) experiments. The RPCs were developed for the MARTA project and were tested on field conditions. These RPCs cover an area of $1.5 \times 1.2\,{m^2}$ and are instrumented with 64 pickup electrodes providing a segmentation better than $20\,$cm. By shielding the detector units with enough slant mass to absorb the electromagnetic component in the air showers, a clean measurement of the muon content is allowed, a concept to be implemented in a next generation of UHECR experiments.
The operation of a ground array detector poses challenging demands, as the RPC must operate remotely under extreme environmental conditions, with limited budgets for power and minimal maintenance.
The RPC, DAQ, High Voltage and monitoring systems are enclosed in an aluminium-sealed case, providing a compact and robust unit suited for outdoor environments, which can be easily deployed and connected. The RPCs developed at LIP-Coimbra are able to operate using a very low gas flux, which allows running them for few years with a small gas reservoir.
Several prototypes have already been built and tested both in the laboratory and outdoors. We report on the most recent tests done in the field that show that the developed RPCs have operated in a stable way for more than 2 years in field conditions.}

\FullConference{35th International Cosmic Ray Conference --- ICRC2017\\
		10--20 July, 2017\\
		Bexco, Busan, Korea}

\begin{document}

\section{Introduction}

Ultra-High Energy Cosmic Rays, the most energetic known particles in the Universe, offer a unique opportunity to study both some of the most enigmatic acceleration processes of our Universe and probe particle physics at a scale currently unaccessible by human-made accelerators.

However, the study of such particles has to be done indirectly by measuring the Extensive Air Showers (EAS) produced by the interaction of the primary particle with the Earth's atmospheric atoms. This requires a knowledge over the shower physical processes, in particular over high-energy hadronic interactions, which we have not achieved yet. In fact, there are multiple evidences that the hadronic component of the cascade is not fully understood~\cite{AugerMu}. These evidences arise mainly from inconsistencies between the shower electromagnetic component and its muonic component. The latter, stems from the decay of charged mesons, being thus intimately related with the description of hadronic interactions.

As such, the accurate and direct measurement of the EAS muon content is crucial for the validation of hadronic interaction models and consequently ensure the correct description of the shower. The direct measurement can be done by a setup like the one proposed for MARTA~\cite{MARTA}. In this setup Resistive Plate Chambers (RPCs) are put below a water-Cherenkov Detector (WCD). The WCD acts like shielding to the electromagnetic shower component allowing the RPCs to detect muons with a high spatial and time resolution~\footnote{The applications of this EAS hybrid detector in terms of inter-calibrations and combined analysis will be discussed in a paper to be published elsewhere.}.
The choice of the RPCs is done not only because of their good time resolution ($< 1\,$ns) and ability to easily be segmented (limited only by the electronics) but also their low cost, as a gaseous detector, makes them extremely appealing for cosmic ray experiments where significantly large areas need to be covered.
However, the introduction of such detectors in a harsh outdoor environment poses additional challenges: these RPCs have to be resilient to environmental effects such as large temperature excursions and humidity; being spread out on the field, maintenance should be low, which implies that these RPCs should have a very low gas flux; it should have a small ageing.
In this communication we present the latest indoor/outdoor results which demonstrates that all the above requirements can be fulfilled. The manuscript is organised as follows: in the next section we describe the detector, after we discuss the low gas flux tests performed at lab and afterwards with describe the outdoor apparatus and present the results for one year of operation. We end with a summary and prospects.

\section{Detector description}

  \begin{figure}
     \centering
	 \includegraphics[align=c, width=.4\textwidth]{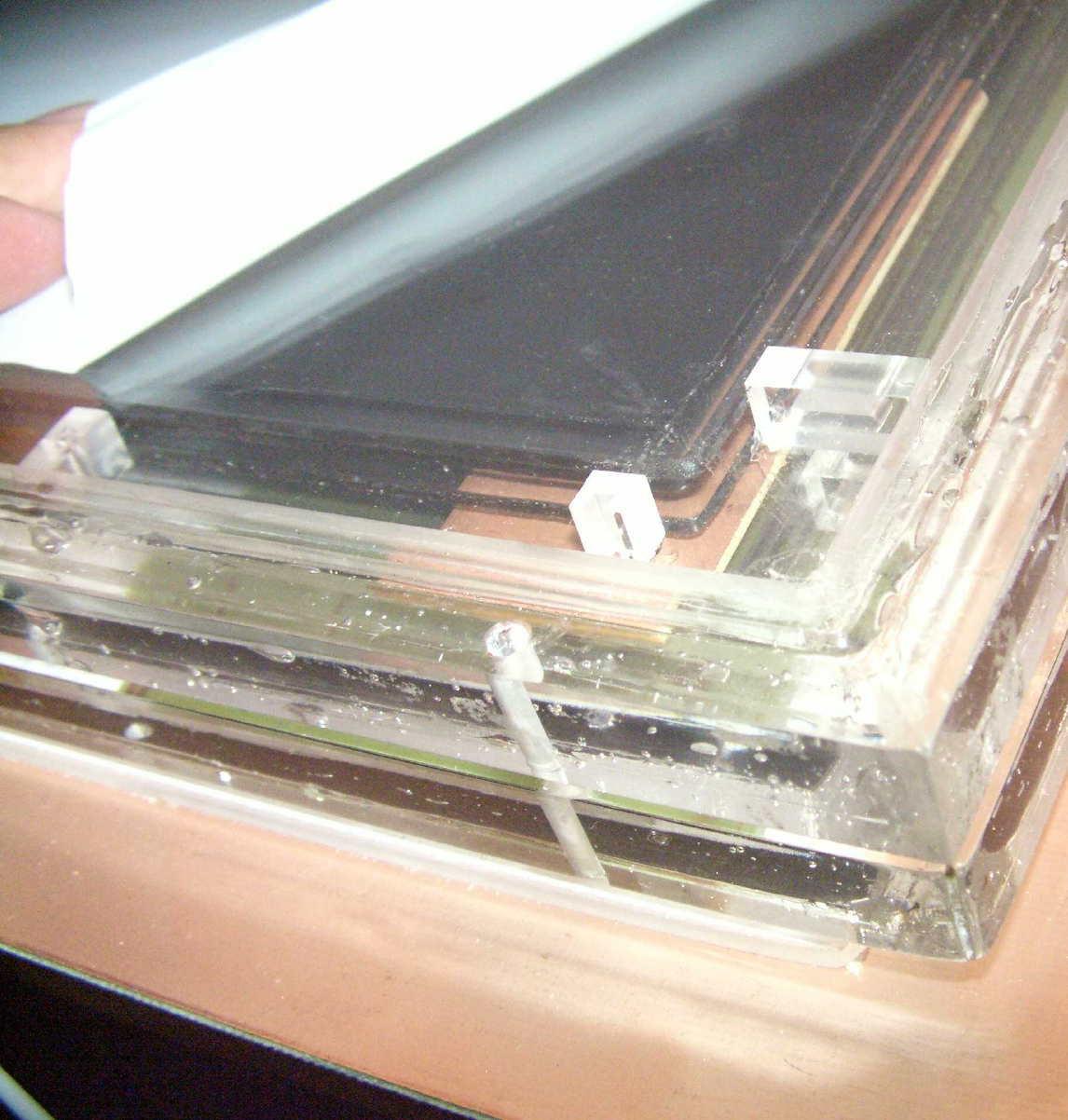}
     \hspace*{5mm}
     \includegraphics[align=c, width=.4\textwidth]{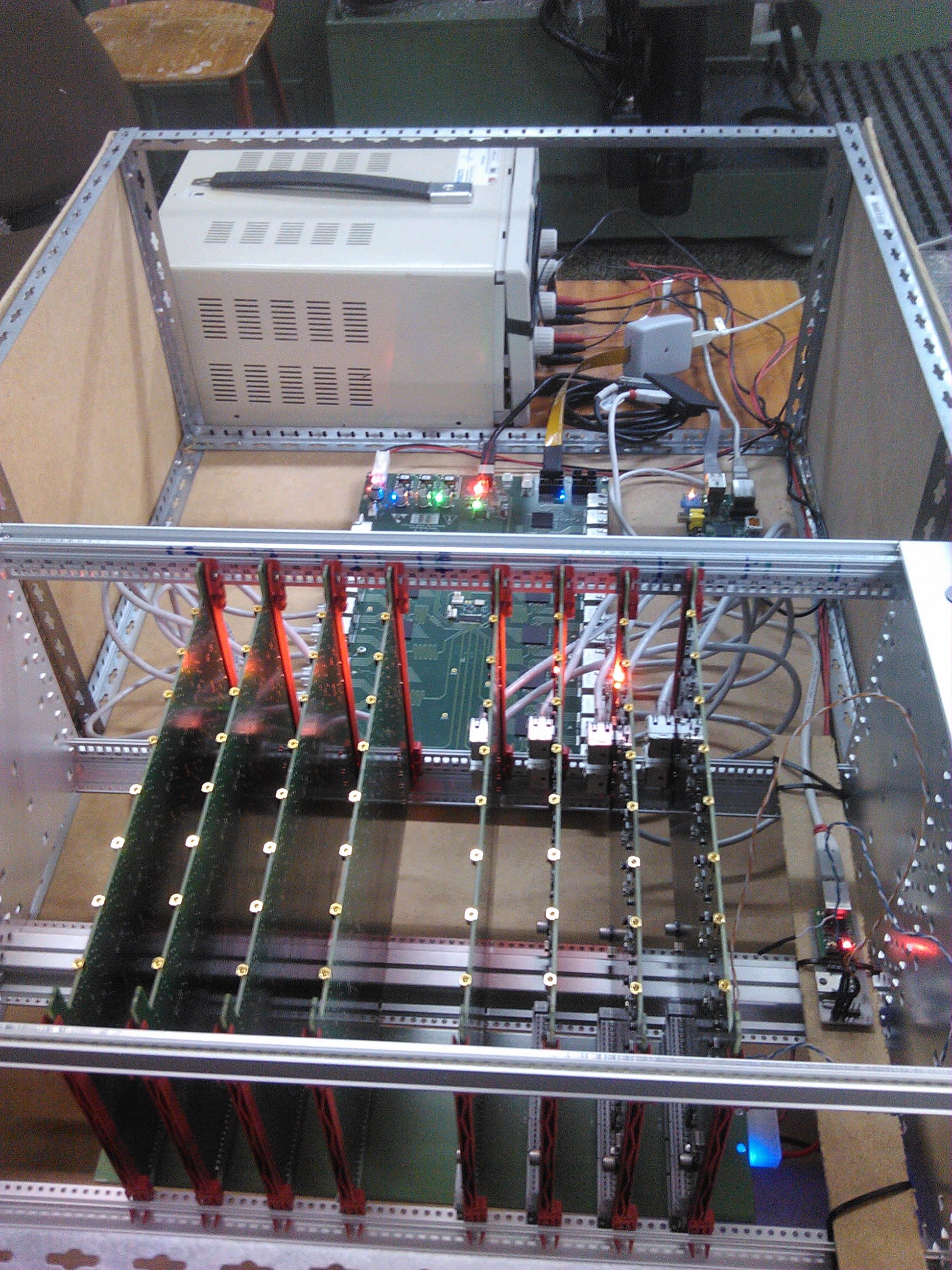}
     \caption{(left) Picture of the acrylic box with the sensitive volume inside. (right) Picture of the DAQ electronics.}
  \label{fig:RPCdetails}
\end{figure}

The Resistive Plate Chamber detector module is composed by a $1200\times1500\times1.9\,{\rm mm^3}$ glass electrodes separated by Nylon monofilaments forming two $1$ mm gas gaps. This module is put inside a permanently closed acrylic box (see figure~\ref{fig:RPCdetails}(left)). The application of a resistive acrylic paint allows one to apply the high voltage on the outer electrodes. %citation HV paint?? Is wrong in LL...
The signal can be read out though a $8\times 8$ pad matrix, where each pad has an area of $180\times 140\,{\rm mm^2}$. The data acquisition (DAQ) electronics, named PREC~\cite{PREC}, is a custom-developed system based on discrete electronics (see figure~\ref{fig:RPCdetails}(right)). Each acquisition channel comprises of a broadband amplifier followed by a programmable comparator. The threshold outputs are sent, via LVDS links, to a purely digital central board. Data remain in this buffer until read by the DAQ computer.

An I2C bus is used to get information about temperature, pressure and relative humidity in the chamber. High voltage (HV) and background currents were monitored by the HV power supply. These parameters are recorded each minute.

\begin{figure}
     \centering
	 \includegraphics[align=c, width=.4\textwidth]{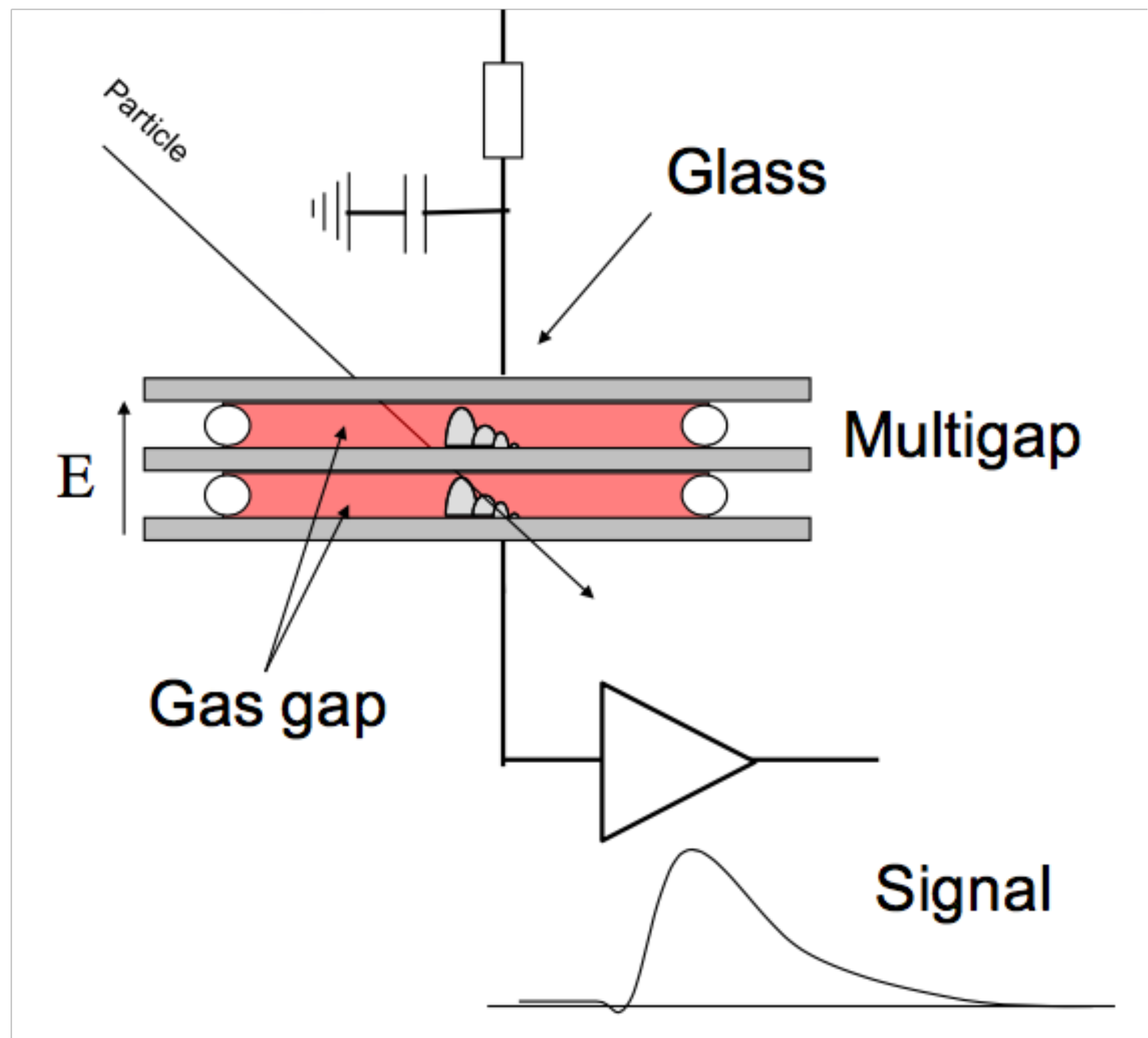}
     \hspace*{5mm}
     \includegraphics[align=c, width=.4\textwidth]{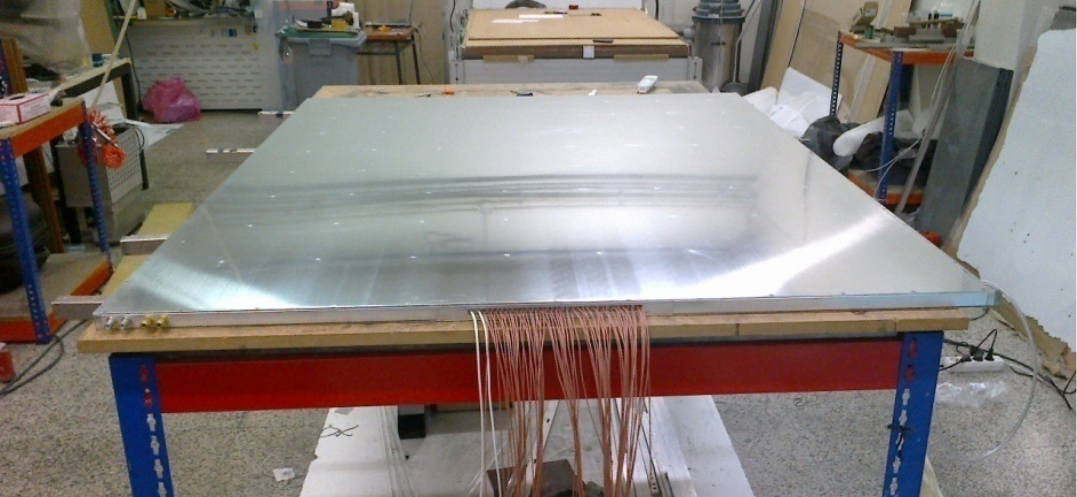}
     \caption{(left) Scheme of the RPC detector operation principle. (right) Image of an assembled RPC.}
  \label{fig:RPC}
\end{figure}

%%PJ: isto nao foi feito aqui... %An important parameter for this study is the gas (pure R-134a) flow rates and the relative humidity at the outputs of the sensitive volume and the aluminium box. This allows to monitor the tightness of both volumes and control the gas flux.
The RPC was designed to be used on a harsh environment and as such both the gaseous volume and the pickup plane are inside a gas-tight aluminium volume (see figure~\ref{fig:RPC}(right)). The HV power supply and the frontend electronics will be located in a DAQ box coupled to the RPC volume.

The RPC is operated in proportional mode which allows it to operate with a high efficiency while minimizing ageing of the RPC due to electric discharges in the gas. 

\section{Laboratory tests}

The gas flux is an important parameter for the RPC operation. As such, although it is desirable to have a gaseous detector with as low gas flux as possible, one needs to be sure that this is not achieved at the cost of losing efficiency. In this section we present the laboratory tests done for several gas fluxes starting at $12\,$cc/min down to $1\,$cc/min.

One of the most important parameters to be monitored and maintained stable is the reduced electric field, $E/N$, which is a function of the RPC gap width, the applied voltage, the temperature and the pressure. This quantity cannot be obtained directly but it can be calculated using the previous mentioned quantities. In this work we followed the approach used in~\cite{LuisRPCs}. 

\begin{figure}
     \centering
	 \includegraphics[align=c, width=1.\textwidth]{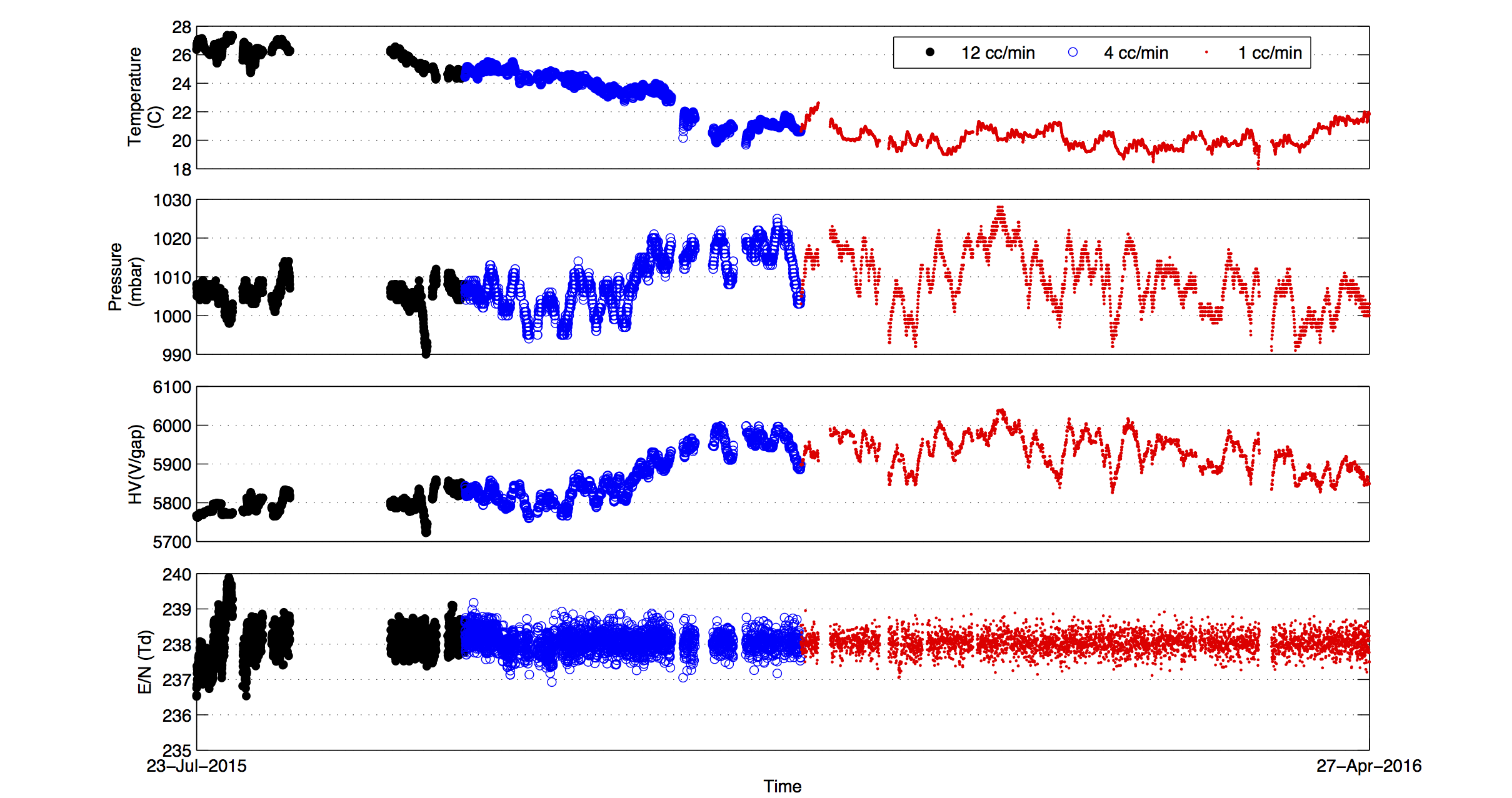}
     \caption{Reduced electric field and the three variables considered for its determination via the automatic adjustment of the applied high voltage. %The arrow indicates the start of the adjustment process. 
     }
  \label{fig:flux1}
\end{figure}

In order to maintain its efficiency with the change of environmental parameters the high voltage (HV) has to be adjusted accordingly. Hence, the HV was adjusted each 15 minutes, using the average data of the previous 15 minutes. This appears to be enough to accommodate pressure and temperature variations as shown in figure~\ref{fig:flux1}. In this figure is presented the evolution of $E/N$ during more than 8 months and for different gas fluxes. From these results, it becomes evident that it is possible to maintain $E/N$ constant for different conditions by adjusting the HV.
%PJ: O E/N nao tem k ver com a qualidade do gas. Eh mesmo so uma conta com HV, distancia, pressao, temperatura....%%it becomes evident that the stability of this quantity is not affected by the gas flow even using scarce flux as low as $1\,$cc/min.

% \begin{figure}
%     \centering
%	 \includegraphics[align=c, width=.8\textwidth]{GasMonitor.pdf}
%     \caption{This is the caption of the figure.}
%  \label{fig1}
%\end{figure}

\begin{figure}
     \centering
	 \includegraphics[align=c, width=\textwidth]{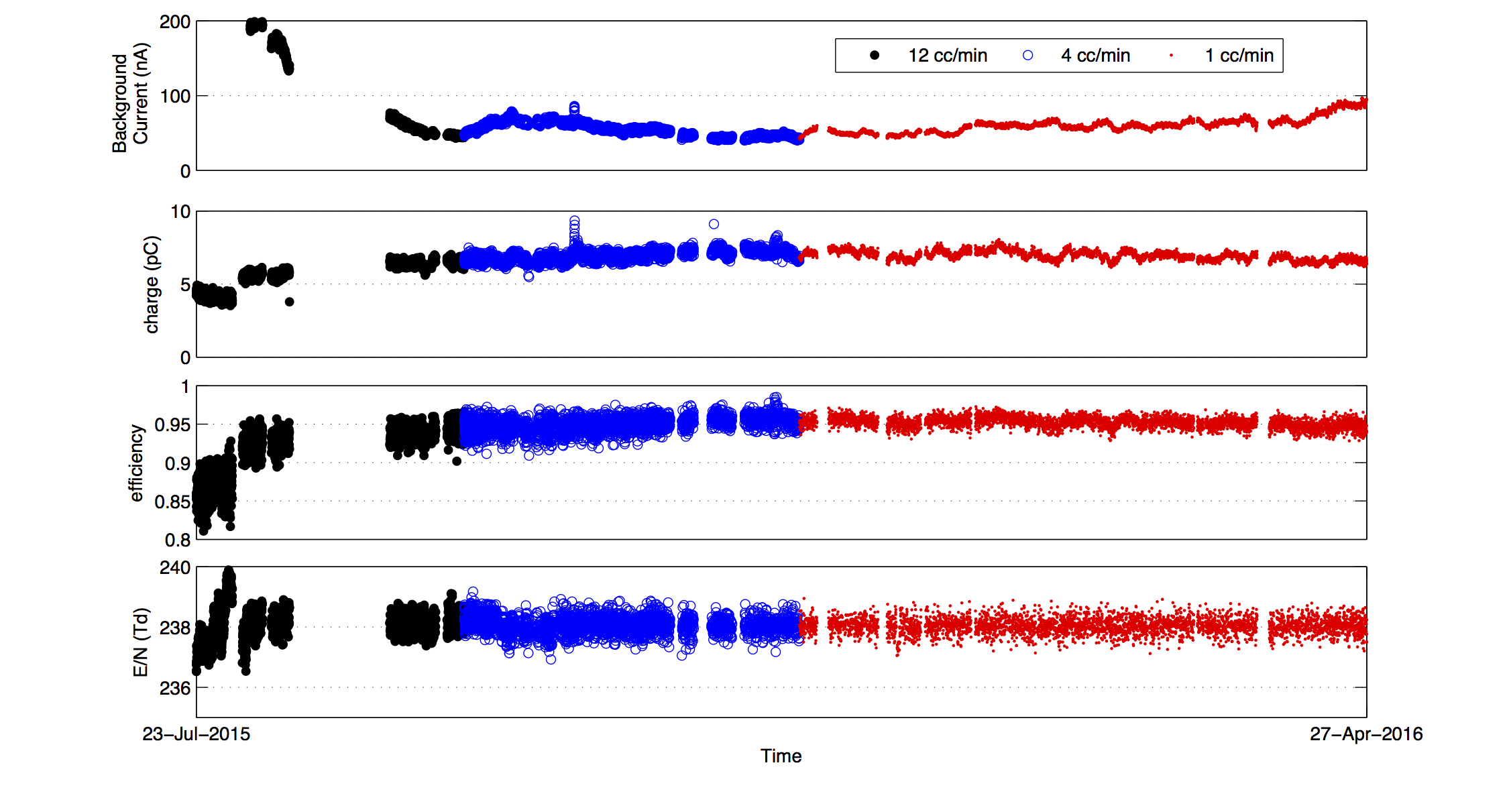}
     \caption{Reduced electric field, efficiency, fast charge and background (operation) current over more than 9 months.}
  \label{fig:flux2}
\end{figure}

The variation of the induced fast charge, efficiency, background current and $E/N$, can be seen in figure~\ref{fig:flux2} over more than 9 months. From these plots is it possible to see that all these quantities vary little over time, even when the gas flux is reduced. Only the background current is not as stable as the other variables. This happens as the current is the sum of various contributions: ionisation currents, leakage currents and the imbalance of the background rate due to temperature variations. It should be however noted that last two contributions will not affect the charge nor efficiency since they are excluded by the trigger definition, and thus have no impact on data analysis of shower events.

\section{Test on the outdoor}

Once the RPCs were proven to fulfil the necessary requirements in the lab, they needed to be tested in the outdoor environment. The test was conducted in the Pampa Amarilla, in the Province of Mendoza, Argentina at the Pierre Auger Observatory site~\cite{Auger}. This plateau has an altitude of $1400\,$m above sea level. The atmospheric conditions are demanding with daily temperature excursions of nearly $30^\circ\,{\rm C}$, minimum temperatures below zero and maximum absolute temperatures exceeding $30^\circ\,{\rm C}$. The RPC has to also be able to endure strong winds and lightning storms as well as high humidity.

\begin{figure}
     \centering
	 \includegraphics[align=c, width=.4\textwidth]{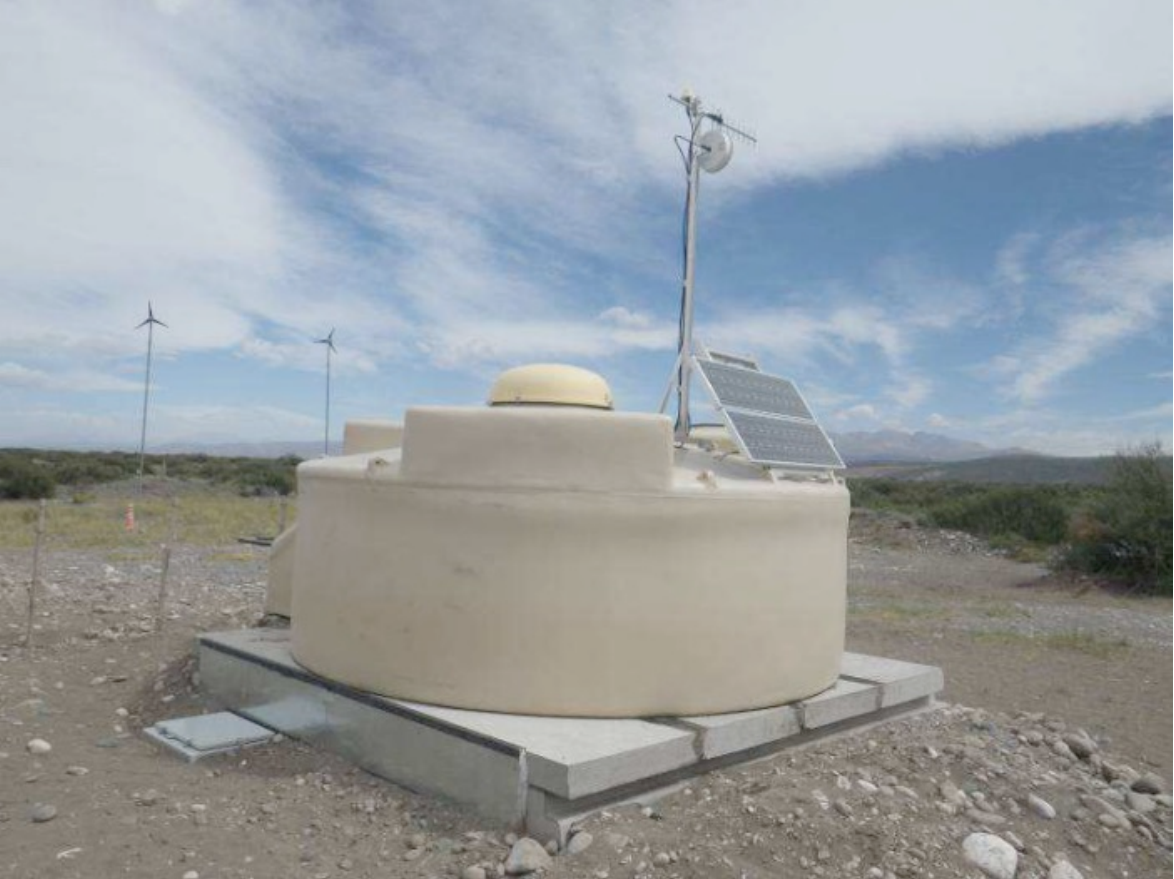}
     \hspace*{5mm}
     \includegraphics[align=c, width=.4\textwidth]{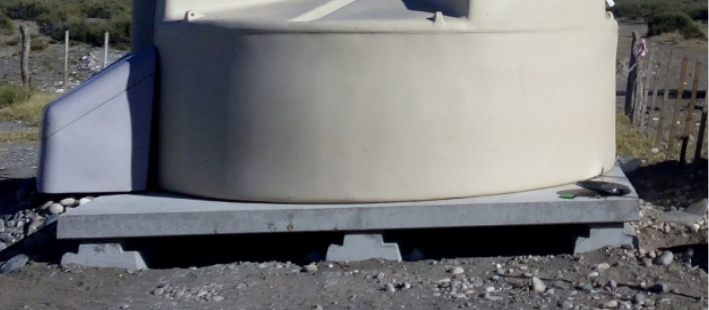}
     \caption{(left) Pierre Auger Observatory water-Cherenkov Detector (\emph{Tierra del Fuego}) with two RPCs below, inside the concrete precast. (right) Detail of the concrete precast while it is open on the side.}
  \label{fig:MARTA}
\end{figure}

\begin{figure}
     \centering
	 \includegraphics[align=c, width=\textwidth]{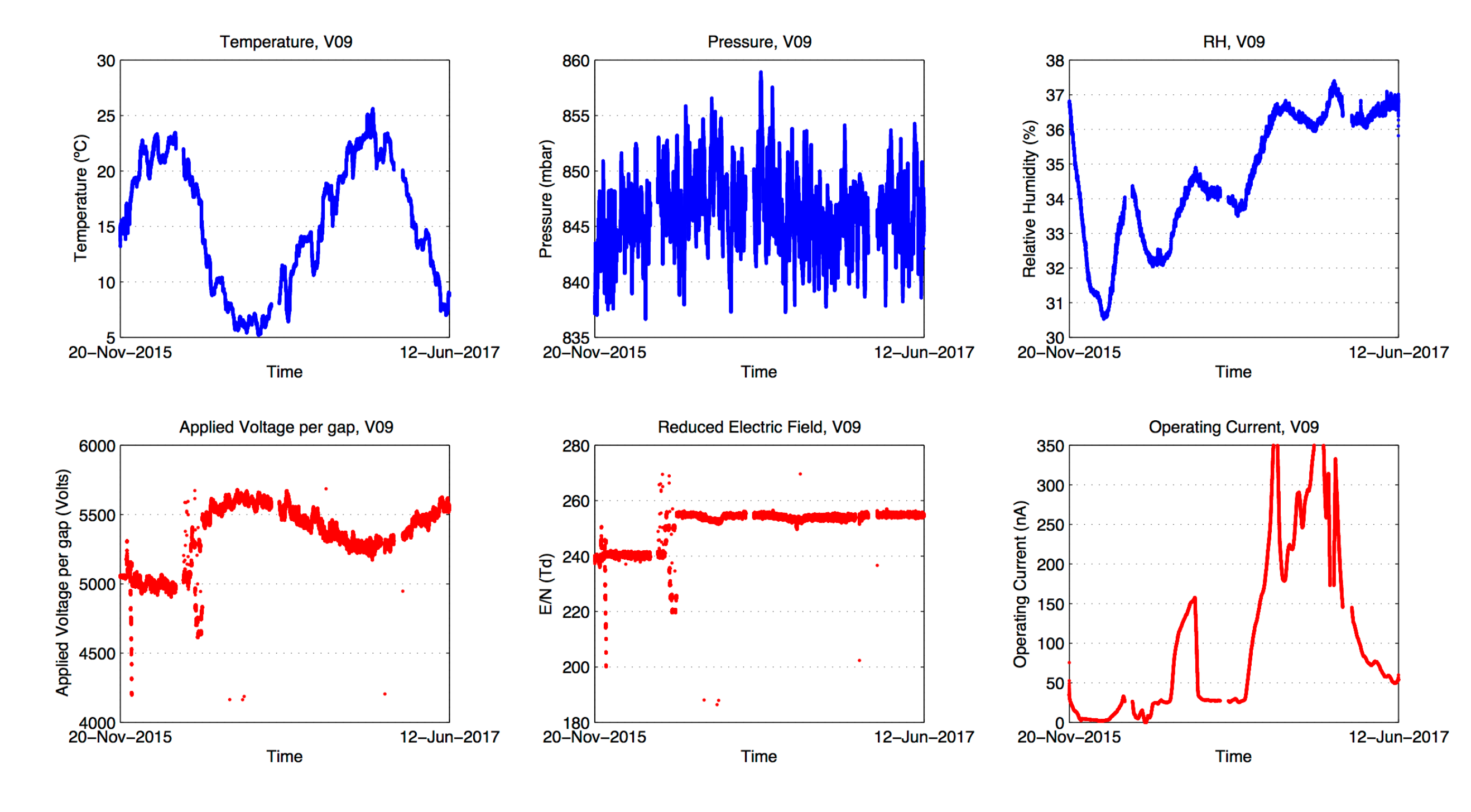}
     \caption{RPC monitored parameters over time (from top to bottom, left to right): inside aluminum box temperature, atmospheric pressure and relative humidity, applied voltage per gap, reduced electric field and operating current.}
  \label{fig:RPCparameters}
\end{figure}

The RPCs were installed in a closed concrete precast that supports a WCD, a tank with 12 tons of purified water. The WCD offers protection against the shower electromagnetic unwanted particles and the whole system has a thermal inertia which attenuates the temperature excursions to seasonal variations of less than $10^\circ\,{\rm C}$, as seen in figure~\ref{fig:RPCparameters}.
In order to measure the efficiency of the RPC, another RPC was put on top of the other spaced by 10 mm. This way we can use the tank and one RPC to define the trigger and measure the efficiency of the other one.
%PJ: Não é garantido que sejam so muoes % to muons
%During this test the RPCs operated with a constant gas flux of (?? nao encontrei este info em nenhum lado).

\begin{figure}
     \centering
	 \includegraphics[align=c, width=.5\textwidth]{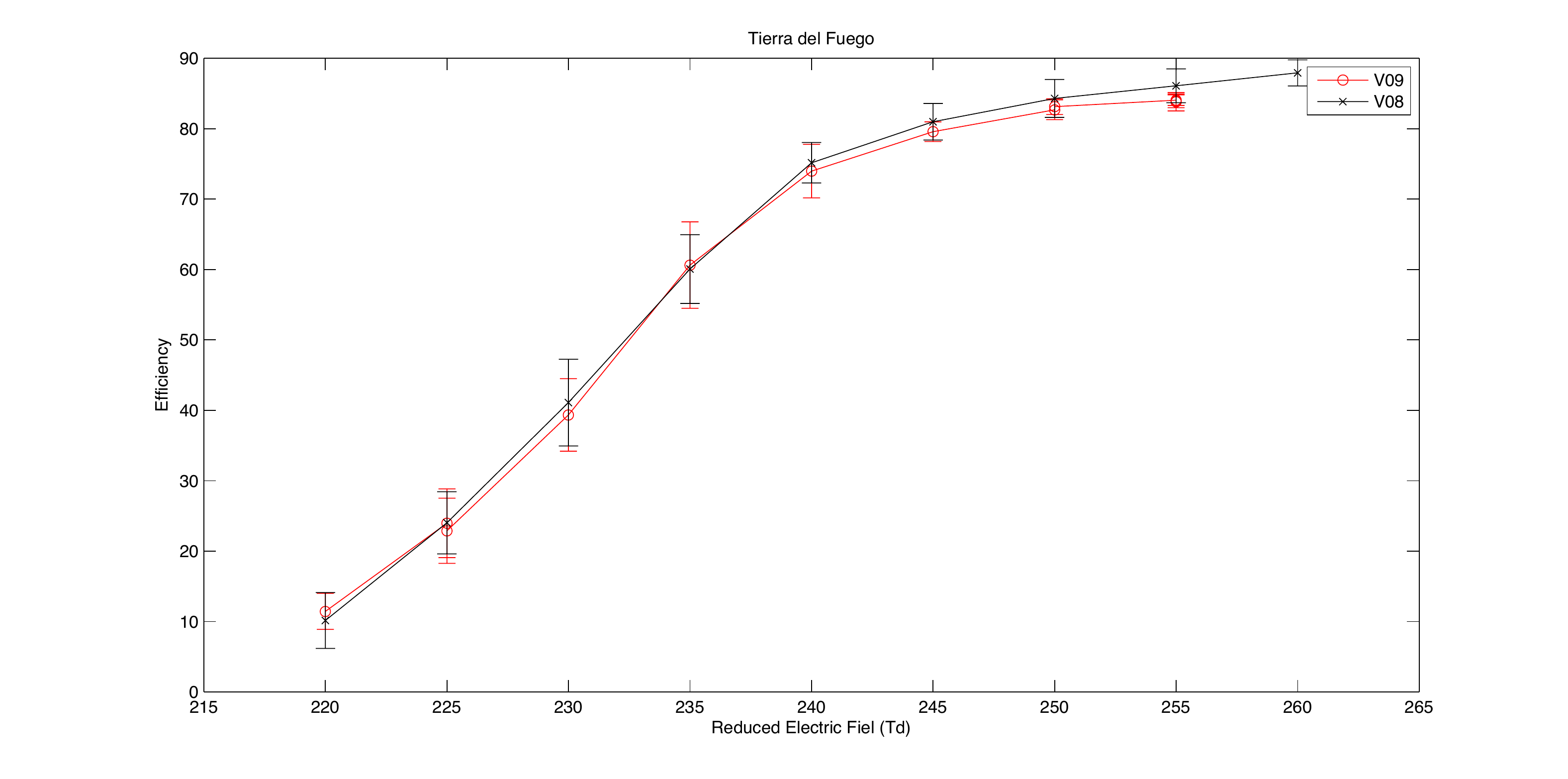}
     \hspace*{5mm}
     \includegraphics[align=c, width=.4\textwidth]{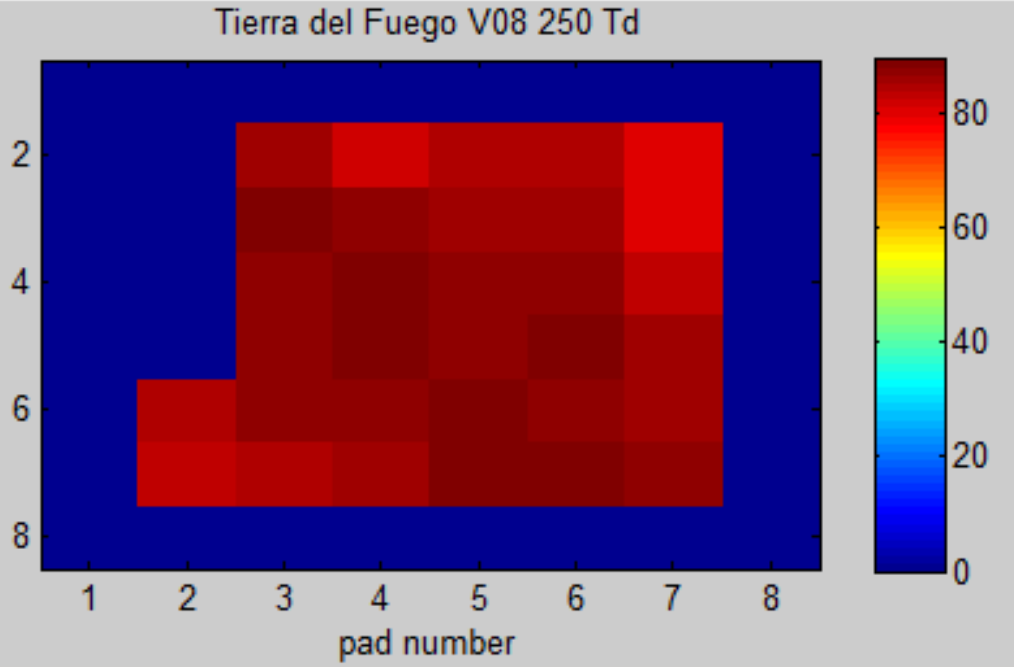}
     \caption{(left) Efficiency as a function of the reduced electric field, $E/N$. (right) Efficiency as a function of the position (uniformity).}
  \label{fig:HV}
\end{figure}

In figure~\ref{fig:HV}(left) it is shown the variation of the RPC efficiency with the reduced electric field. Clearly a plateau can be reached above $\sim 240\,$Td allowing the RPC safe operation. The efficiency uniformity of the RPC is also shown in this figure (right). All the instrumented pads present an efficiency of $\sim 85\%$, showing a good uniformity.

The RPC monitoring parameters can be seen in figure~\ref{fig:RPCparameters} for nearly one year of data acquisition. The most important feature here is that $E/N$ can be maintained constant by adjusting the HV.

\begin{figure}
     \centering
	 \includegraphics[align=c, width=\textwidth]{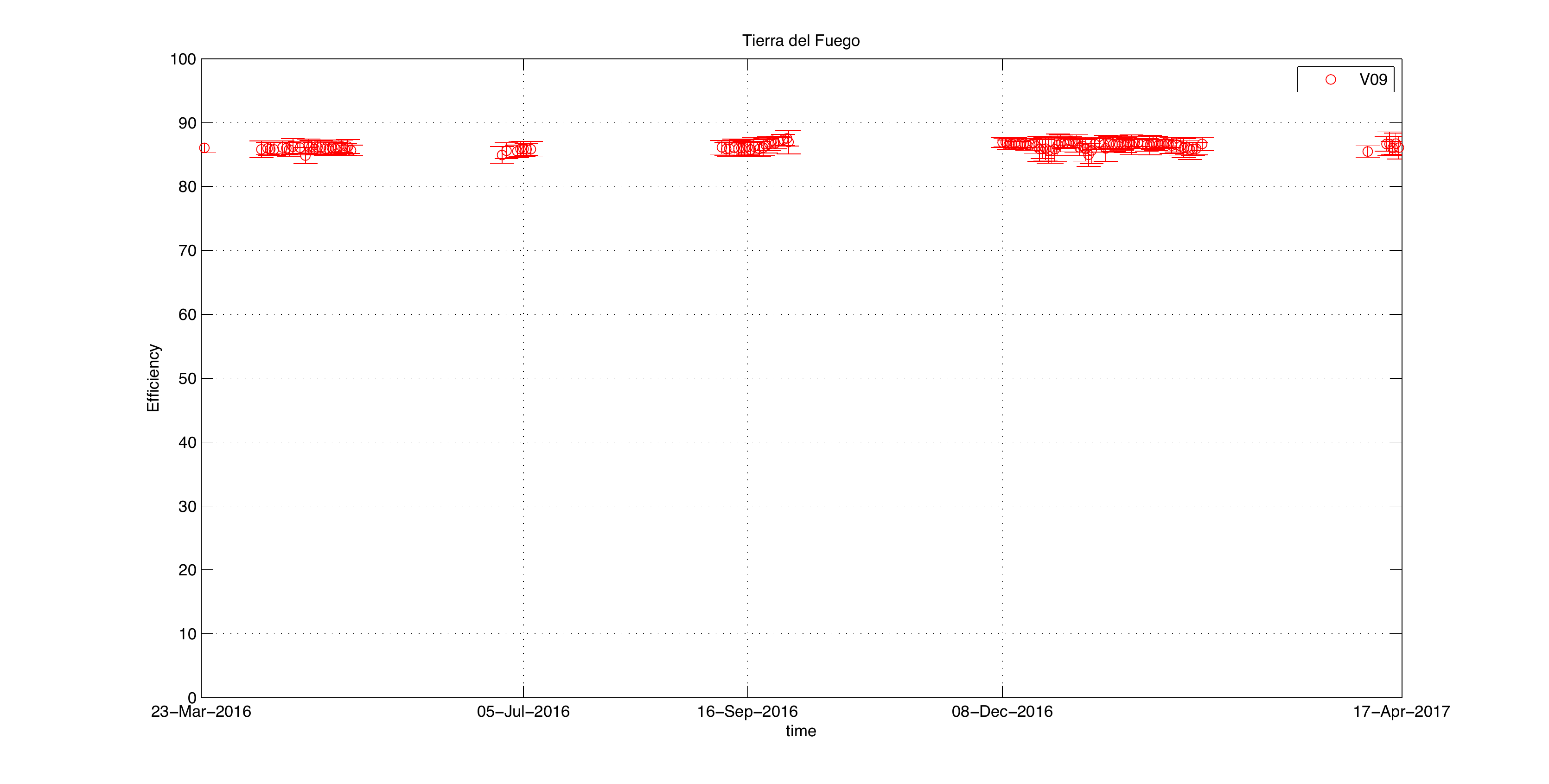}
     \caption{RPC efficiency over nearly 11 months.}
  \label{fig:efficiency}
\end{figure}

This has an immediate impact on the RPC efficiency to muons, shown in figure~\ref{fig:efficiency}. This plot clearly shows that it is possible to operate this RPC in the harsh outdoor environment with constant efficiency, proving its potential for cosmic ray experiments. The observed interruptions of the data acquisition are related with the limited availability of power and communications in the test site. It is worth noting that the important result is not the continuous operation of the RPC but the remarkably stable operation during nearly one year in the field while exposed to different weather seasonal effects. 

\section{Summary and prospects}

The detailed study of Extensive Air Showers requires better detectors, able to operate in the outdoor harsh environment.

Resistive Plate Chambers are a good candidate, due to their low cost and good spatial and time resolution, provided that they can operate stably and with little maintenance.

In this work, we have shown with laboratory and outdoor tests that it is possible to operate the RPCs with gas fluxes as low as
%PJ: Not sure about this. Nao tenho a certeza que se tenha usado um fluxo tao baixo no field.
$1\,$cc/min while maintaining a good efficiency. Moreover, through the adjustment of the gap voltage it is possible to absorb the temperature and pressure variation ensuring stable detection efficiency. This stability was tested during one year in the open field demonstrating the detector resilience to environmental effects.

%PJ: Isto so aparece aqui nas conclusoes? Nao foi introduzido no texto . Penso que se devia cortar pelo menos a parte do hodoscopio
Currently there are about 30 of these RPCs being used and tested in several places in the world. For instance, it is being used at the Pierre Auger Observatory site as a hodoscope to investigate the response of the WCD to muons\cite{pedjorGianni}. Hence, the R\&D of these RPCs for an outdoor operation is expected to continue with more data and further developments.
 
This R\&D is essential for future projects that plan to take advantage of RPCs capabilities. For instance, LATTES~\cite{RubenRICAP}, an array for the detection of (very) high-energy gamma-rays planned to be installed at very high altitude in South America.

\bibliographystyle{jhep}
\bibliography{Biblio}

\end{document}